%Paper: hep-ph/9412210
%From: steif@dirac.ucdavis.edu (Alan Steif)
%Date: Thu, 1 Dec 94 11:43:18 PST
%Date (revised): Mon, 9 Jan 95 10:52:32 PST

%%%%%%%%%%%%%%%%%%  tex macros for preprints, cm version %%%%%%%%%%%%%%
%                     (P. Ginsparg, last updated 9/91)
%                if confused, type `b' in response to query
%
%---------------------------------------------------------------------%
%% site dependent options:
%% \unredoffs and \redoffs define horizontal and vertical offsets
%% respectively for unreduced and reduced modes. \speclscape defines
%% the \special{} call that sets printer to landscape (sideways) mode.
%% from standard set below, leave uncommented as appropriate or redefine
%
%%% next 400dpi
%\def\unredoffs{} \def\redoffs{\voffset=-.31truein\hoffset=-.48truein}
%\def\speclscape{\special{landscape}}
%
%%% apple lw
\def\unredoffs{} \def\redoffs{\voffset=-.31truein\hoffset=-.59truein}
\def\speclscape{\special{ps: landscape}}
%
%%% qms lasergrafix:
%\def\unredoffs{} \def\redoffs{\voffset=-.4truein\hoffset=.125truein}
%\def\speclscape{\special{qms: landscape}}
%
%%% saclay A4 paper:
%\def\unredoffs{\hoffset-.14truein\voffset-.2truein}
%\def\redoffs{\voffset=-.45truein\hoffset=-.21truein}
%\def\speclscape{\special{landscape}}
%
%---------------------------------------------------------------------%
%
\newbox\leftpage \newdimen\fullhsize \newdimen\hstitle \newdimen\hsbody
\tolerance=1000\hfuzz=2pt
\catcode`\@=11 % This allows us to modify PLAIN macros.
\def\bigans{b }
\message{ big or little (b/l)? }\read-1 to\answ
\ifx\answ\bigans\message{(This will come out unreduced.}
\magnification=1200\unredoffs\baselineskip=16pt plus 2pt minus 1pt
\hsbody=\hsize \hstitle=\hsize %take default values for unreduced format
\else\message{(This will be reduced.} \let\l@r=L
\magnification=1000\baselineskip=16pt plus 2pt minus 1pt \vsize=7truein
\redoffs \hstitle=8truein\hsbody=4.75truein\fullhsize=10truein\hsize=\hsbody
\output={\ifnum\pageno=0 %%% This is the HUTP version
  \shipout\vbox{\speclscape{\hsize\fullhsize\makeheadline}
    \hbox to \fullhsize{\hfill\pagebody\hfill}}\advancepageno
  \else
  \almostshipout{\leftline{\vbox{\pagebody\makefootline}}}\advancepageno
  \fi}
\def\almostshipout#1{\if L\l@r \count1=1 \message{[\the\count0.\the\count1]}
      \global\setbox\leftpage=#1 \global\let\l@r=R
 \else \count1=2
  \shipout\vbox{\speclscape{\hsize\fullhsize\makeheadline}
      \hbox to\fullhsize{\box\leftpage\hfil#1}}  \global\let\l@r=L\fi}
\fi
%---------------------------------------------------------------------
%
\newcount\yearltd\yearltd=\year\advance\yearltd by -1900

\def\Title#1#2{\nopagenumbers\abstractfont\hsize=\hstitle\rightline{#1}%
\vskip 1in\centerline{\titlefont #2}\abstractfont\vskip .5in\pageno=0}
\def\Date#1{\vfill\leftline{#1}\tenpoint\supereject\global\hsize=\hsbody%
\footline={\hss\tenrm\folio\hss}}% 	restores pagenumbers
%
%       use following instead of \Date on the preliminary draft,
%       puts date/time on each page in big mode, writes labels in margins

\def\draftmode{\message{ DRAFTMODE }\def\draftdate{{\rm preliminary draft:
\number\month/\number\day/\number\yearltd\ \ \hourmin}}%
\headline={\hfil\draftdate}\writelabels\baselineskip=20pt plus 2pt minus 2pt
 {\count255=\time\divide\count255 by 60 \xdef\hourmin{\number\count255}
  \multiply\count255 by-60\advance\count255 by\time
  \xdef\hourmin{\hourmin:\ifnum\count255<10 0\fi\the\count255}}}
%       use \nolabels to get rid of eqn, ref, and fig labels in draft mode
\def\nolabels{\def\wrlabeL##1{}\def\eqlabeL##1{}\def\reflabeL##1{}}
\def\writelabels{\def\wrlabeL##1{\leavevmode\vadjust{\rlap{\smash%
{\line{{\escapechar=` \hfill\rlap{\sevenrm\hskip.03in\string##1}}}}}}}%
\def\eqlabeL##1{{\escapechar-1\rlap{\sevenrm\hskip.05in\string##1}}}%
\def\reflabeL##1{\noexpand\llap{\noexpand\sevenrm\string\string\string##1}}}
\nolabels
%
% tagged sec numbers
\global\newcount\secno \global\secno=0
\global\newcount\meqno \global\meqno=1
\def\newsec#1{\global\advance\secno by1\message{(\the\secno. #1)}
%\ifx\answ\bigans \vfill\eject \else \bigbreak\bigskip \fi  %if desired
\global\subsecno=0\eqnres@t\noindent{\bf\the\secno. #1}
\writetoca{{\secsym} {#1}}\par\nobreak\medskip\nobreak}
\def\eqnres@t{\xdef\secsym{\the\secno.}\global\meqno=1\bigbreak\bigskip}
\def\sequentialequations{\def\eqnres@t{\bigbreak}}\xdef\secsym{}
\global\newcount\subsecno \global\subsecno=0
\def\subsec#1{\global\advance\subsecno by1\message{(\secsym\the\subsecno. #1)}
\ifnum\lastpenalty>9000\else\bigbreak\fi
\noindent{\it\secsym\the\subsecno. #1}\writetoca{\string\quad
{\secsym\the\subsecno.} {#1}}\par\nobreak\medskip\nobreak}
\def\appendix#1#2{\global\meqno=1\global\subsecno=0\xdef\secsym{\hbox{#1.}}
\bigbreak\bigskip\noindent{\bf Appendix #1. #2}\message{(#1. #2)}
\writetoca{Appendix {#1.} {#2}}\par\nobreak\medskip\nobreak}
%
%       \eqn\label{a+b=c}	gives displayed equation, numbered
%				consecutively within sections.
%     \eqnn and \eqna define labels in advance (of eqalign?)
%
\def\eqnn#1{\xdef #1{(\secsym\the\meqno)}\writedef{#1\leftbracket#1}%
\global\advance\meqno by1\wrlabeL#1}
\def\eqna#1{\xdef #1##1{\hbox{$(\secsym\the\meqno##1)$}}
\writedef{#1\numbersign1\leftbracket#1{\numbersign1}}%
\global\advance\meqno by1\wrlabeL{#1$\{\}$}}
\def\eqn#1#2{\xdef #1{(\secsym\the\meqno)}\writedef{#1\leftbracket#1}%
\global\advance\meqno by1$$#2\eqno#1\eqlabeL#1$$}
%
%			 footnotes
\newskip\footskip\footskip14pt plus 1pt minus 1pt %sets footnote baselineskip
\def\footnotefont{\ninepoint}\def\f@t#1{\footnotefont #1\@foot}
\def\f@@t{\baselineskip\footskip\bgroup\footnotefont\aftergroup\@foot\let\next}
\setbox\strutbox=\hbox{\vrule height9.5pt depth4.5pt width0pt}
\global\newcount\ftno \global\ftno=0
\def\foot{\global\advance\ftno by1\footnote{$^{\the\ftno}$}}
%
%say \footend to put footnotes at end
%will cause problems if \ref used inside \foot, instead use \nref before
\newwrite\ftfile
\def\footend{\def\foot{\global\advance\ftno by1\chardef\wfile=\ftfile
$^{\the\ftno}$\ifnum\ftno=1\immediate\openout\ftfile=foots.tmp\fi%
\immediate\write\ftfile{\noexpand\smallskip%
\noexpand\item{f\the\ftno:\ }\pctsign}\findarg}%
\def\footatend{\vfill\eject\immediate\closeout\ftfile{\parindent=20pt
\centerline{\bf Footnotes}\nobreak\bigskip\input foots.tmp }}}
\def\footatend{}
%
%     \ref\label{text}
% generates a number, assigns it to \label, generates an entry.
% To list the refs on a separate page,  \listrefs
%
\global\newcount\refno \global\refno=1
\newwrite\rfile
\def\ref{[\the\refno]\nref}
\def\nref#1{\xdef#1{[\the\refno]}\writedef{#1\leftbracket#1}%
\ifnum\refno=1\immediate\openout\rfile=refs.tmp\fi
\global\advance\refno by1\chardef\wfile=\rfile\immediate
\write\rfile{\noexpand\item{#1\ }\reflabeL{#1\hskip.31in}\pctsign}\findarg}
%	horrible hack to sidestep tex \write limitation
\def\findarg#1#{\begingroup\obeylines\newlinechar=`\^^M\pass@rg}
{\obeylines\gdef\pass@rg#1{\writ@line\relax #1^^M\hbox{}^^M}%
\gdef\writ@line#1^^M{\expandafter\toks0\expandafter{\striprel@x #1}%
\edef\next{\the\toks0}\ifx\next\em@rk\let\next=\endgroup\else\ifx\next\empty%
\else\immediate\write\wfile{\the\toks0}\fi\let\next=\writ@line\fi\next\relax}}
\def\striprel@x#1{} \def\em@rk{\hbox{}}
\def\lref{\begingroup\obeylines\lr@f}
\def\lr@f#1#2{\gdef#1{\ref#1{#2}}\endgroup\unskip}

\def\addref#1{\immediate\write\rfile{\noexpand\item{}#1}} %now unnecessary
\def\startrefs#1{\immediate\openout\rfile=refs.tmp\refno=#1}
\def\xref{\expandafter\xr@f}\def\xr@f[#1]{#1}
\def\refs#1{\count255=1[\r@fs #1{\hbox{}}]}
\def\r@fs#1{\ifx\und@fined#1\message{reflabel \string#1 is undefined.}%
\nref#1{need to supply reference \string#1.}\fi%
\vphantom{\hphantom{#1}}\edef\next{#1}\ifx\next\em@rk\def\next{}%
\else\ifx\next#1\ifodd\count255\relax\xref#1\count255=0\fi%
\else#1\count255=1\fi\let\next=\r@fs\fi\next}
%
%%%%%%%%%%%%%%%%%%%%%%%%%%%%%%%%%%%%%%%%%%%%%%%%%%%%%%%%%%%%%%%%%%%%%%%%%%
% epsf files
%%%%%%%%%%%%%%%%%%%%%%%%%%%%%%%%%%%%%%%%%%%%%%%%%%%%%%%%%%%%%%%%%%%%%%%%%%
\input epsf
%%%%%%%%%%%%%%%%%%%%%%
% font=8pt

%%%%%%%%%%%%%%%%%%%%%%

%
% this is ugly, but moore insists
\newwrite\ffile\global\newcount\figno \global\figno=1
\def\fig{fig.~\the\figno\nfig}
\def\nfig#1{\xdef#1{fig.~\the\figno}%
\writedef{#1\leftbracket fig.\noexpand~\the\figno}%
\ifnum\figno=1\immediate\openout\ffile=figs.tmp\fi\chardef\wfile=\ffile%
\immediate\write\ffile{\noexpand\medskip\noexpand\item{Fig.\ \the\figno. }
\reflabeL{#1\hskip.55in}\pctsign}\global\advance\figno by1\findarg}
\def\xfig{\expandafter\xf@g}\def\xf@g fig.\penalty\@M\ {}
\def\figs#1{figs.~\f@gs #1{\hbox{}}}
\def\f@gs#1{\edef\next{#1}\ifx\next\em@rk\def\next{}\else
\ifx\next#1\xfig #1\else#1\fi\let\next=\f@gs\fi\next}
\newwrite\lfile
{\escapechar-1\xdef\pctsign{\string\%}\xdef\leftbracket{\string\{}
\xdef\rightbracket{\string\}}\xdef\numbersign{\string\#}}

\def\writestop{\def\writestoppt{\immediate\write\lfile{\string\pageno%
\the\pageno\string\startrefs\leftbracket\the\refno\rightbracket%
\string\def\string\secsym\leftbracket\secsym\rightbracket%
\string\secno\the\secno\string\meqno\the\meqno}\immediate\closeout\lfile}}
\def\writestoppt{}\def\writedef#1{}
\def\seclab#1{\xdef #1{\the\secno}\writedef{#1\leftbracket#1}\wrlabeL{#1=#1}}
\def\subseclab#1{\xdef #1{\secsym\the\subsecno}%
\writedef{#1\leftbracket#1}\wrlabeL{#1=#1}}
\newwrite\tfile \def\writetoca#1{}
\def\leaderfill{\leaders\hbox to 1em{\hss.\hss}\hfill}
%	use this to write file with table of contents
\def\writetoc{\immediate\openout\tfile=toc.tmp
   \def\writetoca##1{{\edef\next{\write\tfile{\noindent ##1
   \string\leaderfill {\noexpand\number\pageno} \par}}\next}}}
%       and this lists table of contents on second pass

%
\catcode`\@=12 % at signs are no longer letters
%
%	Unpleasantness in calling in abstract and title fonts
\edef\tfontsize{\ifx\answ\bigans scaled\magstep3\else scaled\magstep4\fi}
\font\titlerm=cmr7 \tfontsize \font\titlerms=cmr7 \tfontsize
\font\titlermss=cmr7 \tfontsize \font\titlei=cmmi10 \tfontsize
\font\titleis=cmmi7 \tfontsize \font\titleiss=cmmi5 \tfontsize
\font\titlesy=cmsy10 \tfontsize \font\titlesys=cmsy7 \tfontsize
\font\titlesyss=cmsy5 \tfontsize \font\titleit=cmti10 \tfontsize
\skewchar\titlei='177 \skewchar\titleis='177 \skewchar\titleiss='177
\skewchar\titlesy='60 \skewchar\titlesys='60 \skewchar\titlesyss='60
\def\titlefont{\def\rm{\fam0\titlerm}% switch to title font
\textfont0=\titlerm \scriptfont0=\titlerms \scriptscriptfont0=\titlermss
\textfont1=\titlei \scriptfont1=\titleis \scriptscriptfont1=\titleiss
\textfont2=\titlesy \scriptfont2=\titlesys \scriptscriptfont2=\titlesyss
\textfont\itfam=\titleit \def\it{\fam\itfam\titleit}\rm}
 \ifx\answ\bigans\else scaled\magstep1\fi
\ifx\answ\bigans\def\abstractfont{\tenpoint}\else
\font\abssl=cmsl10 scaled \magstep1
\font\absrm=cmr10 scaled\magstep1 \font\absrms=cmr7 scaled\magstep1
\font\absrmss=cmr5 scaled\magstep1 \font\absi=cmmi10 scaled\magstep1
\font\absis=cmmi7 scaled\magstep1 \font\absiss=cmmi5 scaled\magstep1
\font\abssy=cmsy10 scaled\magstep1 \font\abssys=cmsy7 scaled\magstep1
\font\abssyss=cmsy5 scaled\magstep1 \font\absbf=cmbx10 scaled\magstep1
\skewchar\absi='177 \skewchar\absis='177 \skewchar\absiss='177
\skewchar\abssy='60 \skewchar\abssys='60 \skewchar\abssyss='60
\def\abstractfont{\def\rm{\fam0\absrm}% switch to abstract font
\textfont0=\absrm \scriptfont0=\absrms \scriptscriptfont0=\absrmss
\textfont1=\absi \scriptfont1=\absis \scriptscriptfont1=\absiss
\textfont2=\abssy \scriptfont2=\abssys \scriptscriptfont2=\abssyss
\textfont\itfam=\bigit \def\it{\fam\itfam\bigit}\def\footnotefont{\tenpoint}%
\textfont\slfam=\abssl \def\sl{\fam\slfam\abssl}%
\textfont\bffam=\absbf \def\bf{\fam\bffam\absbf}\rm}\fi
\def\tenpoint{\def\rm{\fam0\tenrm}% switch back to 10-point type
\textfont0=\tenrm \scriptfont0=\sevenrm \scriptscriptfont0=\fiverm
\textfont1=\teni  \scriptfont1=\seveni  \scriptscriptfont1=\fivei
\textfont2=\tensy \scriptfont2=\sevensy \scriptscriptfont2=\fivesy
\textfont\itfam=\tenit \def\it{\fam\itfam\tenit}\def\footnotefont{\ninepoint}%
\textfont\bffam=\tenbf \def\bf{\fam\bffam\tenbf}\def\sl{\fam\slfam\tensl}\rm}
\font\ninerm=cmr9 \font\sixrm=cmr6 \font\ninei=cmmi9 \font\sixi=cmmi6
\font\ninesy=cmsy9 \font\sixsy=cmsy6 \font\ninebf=cmbx9
\font\nineit=cmti9 \font\ninesl=cmsl9 \skewchar\ninei='177
\skewchar\sixi='177 \skewchar\ninesy='60 \skewchar\sixsy='60
\def\ninepoint{\def\rm{\fam0\ninerm}% switch to footnote font
\textfont0=\ninerm \scriptfont0=\sixrm \scriptscriptfont0=\fiverm
\textfont1=\ninei \scriptfont1=\sixi \scriptscriptfont1=\fivei
\textfont2=\ninesy \scriptfont2=\sixsy \scriptscriptfont2=\fivesy
\textfont\itfam=\ninei \def\it{\fam\itfam\nineit}\def\sl{\fam\slfam\ninesl}%
\textfont\bffam=\ninebf \def\bf{\fam\bffam\ninebf}\rm}
%
%---------------------------------------------------------------------
%

\hyphenation{anom-aly anom-alies coun-ter-term coun-ter-terms}
\def\inv{^{\raise.15ex\hbox{${\scriptscriptstyle -}$}\kern-.05em 1}}

\def\Dsl{\,\raise.15ex\hbox{/}\mkern-13.5mu D} %this one can be subscripted
\def\dsl{\raise.15ex\hbox{/}\kern-.57em\partial}

\font\bigit=cmti10 scaled \magstep1
 %pound sterling
\def\lspace{\ifx\answ\bigans{}\else\qquad\fi}
\def\lbspace{\ifx\answ\bigans{}\else\hskip-.2in\fi} % $$\lbspace...$$
\def\boxeqn#1{\vcenter{\vbox{\hrule\hbox{\vrule\kern3pt\vbox{\kern3pt
	\hbox{${\displaystyle #1}$}\kern3pt}\kern3pt\vrule}\hrule}}}
\def\mbox#1#2{\vcenter{\hrule \hbox{\vrule height#2in
		\kern#1in \vrule} \hrule}}  %e.g. \mbox{.1}{.1}
%	matters of taste
%\def\tilde{\widetilde} \def\bar{\overline} \def\hat{\widehat}
%
% some sample definitions
  %     curly letters

\def\darr#1{\raise1.5ex\hbox{$\leftrightarrow$}\mkern-16.5mu #1}
 %pound sterling

 %puts a small half in a displayed eqn
\def\roughly#1{\raise.3ex\hbox{$#1$\kern-.75em\lower1ex\hbox{$\sim$}}}
 %\magnification \magstep1

\rightline{DAMTP/R94/53}
\Title {UCD-PHY-94-41}
{{\vbox {\centerline{Sphalerons and  Conformally Compactified Minkowski
Spacetime   }}}}

\bigskip
\centerline{G.W. Gibbons}
  \centerline{\it Department of Applied Mathematics and Theoretical
Physics}
\centerline {\it Cambridge University}
\centerline{\it Silver St. }
 \centerline{\it Cambridge,   CB3 9EW }
\centerline {\it United Kingdom }
\centerline{\it gwg1@amtp.cam.ac.uk}
\vskip .2in
 \centerline{Alan R. Steif}
 \centerline{\it Department of
Physics}
\centerline {\it  University of California }
  \centerline{\it Davis, CA  95616}
\centerline {\it United States }
 \centerline{\it steif@dirac.ucdavis.edu}
 \vskip .2in
 \vskip .2in

\noindent
ABSTRACT:
Solutions to the Yang-Mills field equations  which describe exploding or
imploding shells of gauge field and which may serve as approximations to the
exploding sphaleron are discussed. The solutions are conformally related to
$SO(2)\times SO(4)$ invariant Yang-Mills solutions. The behavior of fermions in
the gauge field background and higher gauge groups are also considered.

\Date{}
\vskip 1cm
{\newsec    {Introduction}}

The Yang-Mills field equations in $3+1$ dimensions fail to admit particle-like
solutions. Due to the repulsive stresses, it is energetically favourable
for the gauge field  to expand. In
this paper, we discuss exact solutions
to the Yang-Mills field equations describing
spherical pulses of Yang-Mills field exploding or imploding
at the speed of light. These solutions may  serve as
approximations to the exploding or imploding sphaleron solutions.
 %{\ref\us1{ G. W. Gibbons and A. R. Steif {\it Phys. Lett. B} {\bf  314} 13
%%%(1993).}}{{\ref\us2{
% {\ref\us{{ G. W. Gibbons  and  A. R. Steif {\it Phys. Lett. B} {\bf 320}
%%%(1994) 245.}}}
% {\ref\glav{
%G. Lavrelashvili, ``Fermions in the Background of Dilatonic Sphalerons'',
%MPI-PhT/94/65, hep-th/9410178.}}
A sphaleron is an unstable static solution of the classical
Yang-Mills equations coupled to an attractive matter field, typically
Higgs {\ref\nm {N. Manton,  {\it Phys. Rev. D} {\bf 28} (1983)
 2019; F. Klinkhamer and N. Manton,  {\it Phys. Rev. D} {\bf 30} (1984)
2212.}}, dilaton {\ref\lmd{G. Lavrelashvili and D. Maison, {\it Phys. Lett. B}
{\bf 295}
(1992) 67; P. Bizon, {\it Phys. Rev. D} {\bf 47} (1993) 1656.}}, or
gravitational {\ref\bm{
R. Bartnik and J. McKinnon, {\it Phys. Rev. Lett.} {\bf 61} (1988) 141; D.
Gal'tsov and M. Volkov, {\it Phys. Lett. } {\bf B273} (1991) 255; D. Sudarsky
and R. Wald, {\it Phys. Rev. D} {\bf 46} (1992) 1453.}},
 which owes its existence to the fact that the configuration space $ \cal C$ of
the Yang-Mills field is not simply connected.
In fact,  $ \cal C=A/G $, where $ \cal A$ is the affine space of Yang-Mills
connections on a topologically trivial Cauchy surface $\Sigma$ which fall off
suitably at infinity, and $\cal  G$ is the set of gauge transformations which
tend to the identity at infinity. The space $ \cal G$ falls into  disconnected
components corresponding to elements of
$\pi _0 ({\cal G}) = \pi _ 3 (G)$,
where $G$ is the gauge group. Thus, the
fundamental group
$\pi _1 ( {\cal C}) = \pi _1( { \cal A}) / \pi _0 ({\cal G})$
is non-trivial in general.

The energy functional has a maximum
around each closed loop $\gamma $ in $ \cal C$. A sphaleron configuration may
be  obtained by minimizing this maximum among
all closed loops in a non-trivial homotopy class. In the case that the gauge
group $G= SU(2)$, $\pi _0({\cal G}) = Z$, the eponymous sphaleron corresponds
to the fundamental generator of $\pi _1 ({\cal C})$.

  An alternative description is obtained by passing to the universal
covering space $\tilde {\cal C}$ of the configuration space $ {\cal C}$. The
vacuum or least energy configuration in ${\cal  C}$ lifts to many vacua in
$\tilde { \cal C }$
labelled by elements of $\pi _ 0 ({\cal G})$.
The vacua on $\tilde {\cal C} $ are said to be related by large gauge
transformations, i.e. by gauge transformations which are not connected
to the identity. A closed loop $\gamma $ in $ \cal C$ lifts to a curve
$\tilde{ \gamma} $
joining adjacent vacua. One may think of a sphaleron as a configuration
at the top of a potential barrier separating adjacent vacua.

 The Chern-Simons expression
$$
N_{CS} = {e \over {8 \pi ^2}} \int _ \Sigma {\rm Tr} ( A \wedge dA - {{ 2ie}
\over 3} A \wedge A \wedge A )
\eqno (1.1)
$$
defines a real valued  quantity on $\tilde {\cal C}$ which changes by an
integer as one passes between different vacua related by large gauge
transformations, i.e. by elements of $\pi _0({\cal G})$, but is unaltered
by small gauge
transformations, i.e. those connected to the identity. Thus (1.1) defines
an invariant on ${\cal C}$ taking values in $R/Z$. One may regard $N_{CS}$ as
the line integral of a one-form on ${\cal  C}$ which is closed but not exact.

  As one passes around a closed loop $\gamma$ in $ {\cal C}$ the
quantity $N_{CS}$ changes by unity. If $\gamma$ is taken to be
the steepest descent path,
the energy functional is, for the theories under consideration,
 symmetric about the mid-point. The quantity $N_{CS}$ takes the
value ${1 \over 2}$ for the static sphaleron solution.

  The static sphaleron configuration can only exist by
virtue of the coupling of the Yang-Mills field to some other matter
field such as a Higgs, dilaton or gravitational field. There is no
static finite energy solution of pure Yang-Mills theory in flat
spacetime, essentially because of the repulsive stresses exerted
 by the Yang-Mills fields. These can be counterbalanced by attractive forces
due to scalars or gravity, but any such equilibrium is unstable.
 Any slight disturbance will cause the sphaleron to explode outwards
or to collapse inwards
 {\ref\zhou{ Z. Zhou, {\it Helv. Physica Acta.} {\bf 65}  (1992) 767.}}.
 In the absence of gravity one expects the inwards
implosion to be reversed as the inwardly falling spatial configuration
passes through itself and re-expands outwards. For gravitating sphalerons
the situation is more complicated and black hole formation is also possible.

 The time-dependent collapse or explosion of sphalerons is
accompanied by changes in the fermion number of the quantum state of
spinor  fields coupled to the sphaleron. If collapse to a black hole does
not take  place, then  the total net change in fermion number may be
evaluated using the theory of  the chiral anomaly and the Atiyah-Singer
index theorem. An alternative procedure is to use the method of Bogoliubov
transformations. The anomalous production of fermions is then a consequence
of ``strong''  Bogoliubov transformations. If black hole formation
occurs, then there are further complications due to Hawking radiation. In
order to  follow in detail what will happen requires a good understanding of
solutions of the Dirac equation in the the presence of the sphaleron.

  A closely related situation in which more detailed information is desirable
is that envisaged by Christ {\ref\christ{N. Christ, {\it Phys. Rev. D} {\bf21}
(1980) 1591.}} in which baryon number violation might occur during the
collision of gluons or in gluon jets. The purpose of this paper is to provide
such information in the special case that one can ignore
all bosonic fields except the Yang-Mills field. The resulting system of
equations is then conformally invariant,  and we are able to re-interpret
earlier work on classical Yang-Mills theory and conformal invariance
{\ref\conf{
 M. Flato, J. Simon and  D. Sternheimer, {\sl Ann. Phys.} {\bf 61}   (1970)
78.}}
{\ref\others{
  J. Cervero, L. Jacobs and C. R. Nohl, {\sl Phys. Lett. } {\bf B 69} (1977)
351;
 M. Luscher, {\sl Phys. Lett.} {\bf B 70}  (1977) 321;
  B. M. Schecter, {\sl Phys. Rev.} {\bf D 16}  (1977) 3015;
  W. Bernreuther, {\sl Phys. Rev.} {\bf D 16} (1977) 3609;
  J. Harnad and  L. Vinet, {\sl Phys. Lett.} {\bf B76} (1978)  589;
 J. Harnad,  S. Shnider and L. Vinet, {\sl J. Math. Phys.} {\bf 20}   (1979)
931.}}
. We shall argue that the $SO(2) \times SO(4)$ symmetric solution of De Alfaro,
Fubini and Furlan {\ref\meron{ V. De Alfaro, S. Fubini and
 G. Furlan {\it Phys. Lett.  }
{\bf B65}   (1976) 163;
  V. De Alfaro, S. Fubini and  G. Furlan {\it Phys. Lett.  }
{\bf B72}  (1977) 203.}} may serve as a useful model for the behaviour of the
exploding (or by time-reversal ) imploding sphaleron.

 % There are points of similarity and points of difference with the case of the
%%%collapse of cosmic textures discussed by Turok and Spergel [].

\newsec{ Yang-Mills Solutions}

We now review the construction of solutions to the Yang-Mills field equations
describing imploding and exploding shells of gauge field. One begins with known
 $ SO(2) \times SO(4)$ invariant solutions to the field equations on  $S^1
\times S^3$
with  $S^1$  the time direction periodically identified.      We then use the
conformal compactification of  Minkowski space  to $S^1 \times S^3$
to pull back the Yang-Mills solution. Taking advantage of the conformal
invariance of the Yang-Mills equations, one obtains a solution on Minkowski
space that
  describes
shells of Yang-Mills field imploding and then exploding
at the speed of light.
%Similarly, we can study the behavior of fermions in the Yang-Mills background.
%%%There is a fermion zero mode on the $S^1 \times S^3$
% background. Mapping this solution to Minkowski space and using the conformal
%%%invariance of the Dirac equation, one obtains a solution in  Minkowski space
%describing a fermion which is peaked on the Yang-Mills pulse.

{\subsec {   $SO(4)$ and  $ SO(2) \times SO(4)$ invariant Yang-Mills
Solutions}}

%In this section we wish to argue that by taking the $SO(2) \times SO(4)$
%%%invariant solutions [] of $SU(2)$ Yang-Mills theory on $M^{\#}$ and
%%conformally %transforming to Minkowski spacetime $M$ we obtain a model for a
%%sphaleron shell collapsing towards the origin and then expanding outward
%again.

 The $SO(2) \times SO(4)$ solutions have been rather well studied
and we shall not therefore give a detailed derivation. Some information is
given in our earlier paper  {\ref\us{{G.  Gibbons  and  A.  Steif, {\it Phys.
Lett. } {\bf B320} (1994) 245.}}}.
 Consider the group $U(2)$ expressed as $U(1) \times SU(2)/ \pm 1$:
\eqn\U{
U =
\pmatrix {X^5+iX^0 & 0 \cr
          0 & X^5 + i X^0 \cr
}
g,\quad g\equiv\pmatrix { X^4 +iX^3 & iX^1 + X^2 \cr
           iX^1 - X^2 & X^4 -iX^3 \cr
}
}
with
\eqn\uone{
|X^5+iX^0 | ^2 =1
}
and
\eqn\sthree{
|X^4 +iX^3 |^2 + |X^1 +iX^2| ^2 =1,
}
and with the points $X^A \equiv(X^0,X^1,X^2, X^3,X^4,X^5)$
and $-X^A$   identified. Defining
$e^{i\eta} = X^5 + iX^0$, we may consider
%$\tilde M ^ {\#}\equiv
   the $SO(4)$ invariant Yang-Mills connection
\eqn\conn{
 A = {i \over e} f(\eta) g^{-1} dg
}
on $ S^1 \times S^3$ with $e$ the gauge coupling constant.
The Chern-Simons number $N_{CS}$ is given by
\eqn\cs{
N_{CS} = 3 f^2 ( 1 - {2 \over 3} f) .
}
If $f=0$ or $f=1$ the connection {\conn} is flat. These configurations
correspond to two adjacent vacua on $\tilde {\cal C}$. The configuration $f=1$
has $N_{CS}=1$ since the map $S^3 \rightarrow SU(2)$ given by $g$ has degree
one. The Yang-Mills equations will be satisfied if
\eqn\redeq{
{{d ^2 f } \over { d \eta ^2} } - 4 (1-2 f) (f^2-f) =0 .
}
The $SO(2) \times SO(4)$ invariant solution  is independent of $\eta$ and has
$f={ 1 \over 2}$ and therefore
$N_{CS}={1 \over 2}$.   As pointed out in {\us} any solution of the form
{\conn}  with {\redeq}  has the energy momentum tensor of a perfect radiation
fluid
with
\eqn\energydensity{
\rho = {{12E} \over {e^2}} .
}
$E$ being the conserved energy of the effective particle with position $f$
 described by equation {\redeq}. The $SO(2)$ invariant solution $f={ 1 \over
2}$ has $E= { 1 \over 8}$.

{\subsec{  Compactified Minkowski Spacetime}}

We now review the conformal compactification of Minkowski space to
$S^1\times S^3$.
Conformal compactification is the Lorentzian analog of stereographic
projection. Stereographic projection and conformal compactification in any
dimension  can be described in a unified way by considering the light cone in a
space of two higher dimensions.  The two spaces  mapped to one another are then
hyperplanar cross-sections of the light cone, with the mapping   achieved by
projection along the null generators of the light cone. The mapping is
necessarily conformal.

 Consider now the particular case of the conformal compactification of
Minkowski spacetime, $M\equiv M^{3,1}$.
The space of two higher dimensions is the
 six-dimensional flat space, $M^{2,4}$,
\eqn\mmetric{
ds^2 = -(dX^0)^2 - (dX^5)^2 + (dX^1)^2 + (dX^2)^2 + (dX^3)^2 + (dX^4)^2
}
and the  light cone, $K$, centered at the origin is given by
\eqn\nullcone{
 ( X^0)^2 + ( X^5)^2 - ( X^1)^2 - ( X^2)^2 - ( X^3)^2 - ( X^4)^2=0 .
}
Four-dimensional  Minkowski spacetime, $M ,$ is the intersection of $K$ with
the null hyperplane, $X^5=-X^4+1$, where $(X^0 , X^i)\equiv (t, \vec x) \in M$
and $X^4 = (t^2 -r^2 +1)/2$,  $X^5 = (-t^2 +r^2 +1)/2$, $(r \equiv |\vec x |)$.
Similarly, the intersection of $K$ with the $5$-sphere,
\eqn\fivesphere{
 ( X^0)^2 + ( X^5)^2 + ( X^1)^2 + ( X^2)^2 + ( X^3)^2 + ( X^4)^2= 1
}  is
the periodically identified Einstein static universe (ESU)
\eqn\esu{
S^1 \times S^3 = \biggl \{( X^0)^2 + ( X^5)^2 = 1,\;
 ( X^1)^2 + ( X^2)^2 + ( X^3)^2 + ( X^4)^2 =1 \biggr \} .
} On $S^1\times S^3$  define coordinates
$(\eta, \chi, \theta, \phi )$
 \eqn\coords{\eqalign{
X^0 +i X^5 &= \exp i \eta\cr
 X^1 +iX^2 &= \sin\chi \sin \theta \exp i \phi\cr
   X^3 = \sin& \chi \cos \theta \cr
X^4 = \cos& \chi .\cr
}}
The mapping from  ESU to $M$ is obtained by projection along the null
generators of $K$
 and takes
the form
\eqn\map{\eqalign{
v\equiv t+r& = \tan { {\eta + \chi} \over 2 } \cr
 u\equiv t-r& = \tan {{ \eta - \chi } \over 2}.\cr
}}
The mapping is $2 \rightarrow 1$ since $X^{A}$ and $-X^{A}$ (or $(\eta, \chi,
\theta, \phi)$ and $(\eta + \pi, \pi - \chi, \pi - \theta, \phi + \pi)$) map to
the same point of $M$. Identifying  these points and defining $M^{\#}\equiv S^1
\times S^3 / \pm 1 $, we obtain a $1-1$ conformal compactification of $M$ onto
$M^{\#}$.  We note that  the antipodal maps on odd spheres are connected to the
identity diffeomorphism so that {\sl topologically} $M^{\#} \equiv S^1 \times
S^3$.
The mapping takes the flat metric on $M$  to the metric
\eqn\metric{\eqalign{
ds^2 &= \Omega ^{-2} \biggl ( -d \eta ^2 + d \chi ^2 + \sin ^2 \chi
( d \theta ^2 + \sin ^2 \theta d \phi ^2) \biggr )\cr
 \Omega = \cos \eta + \cos \chi
& = {2 \over {\sqrt {(1+u^2)(1+v^2)}}}
= {2 \over {\sqrt
{(1+(t-r)^2)(1+(t+r)^2)}}}\cr
  }}
which is indeed conformal to the standard product metric on ESU.
Future  and past null infinity  ${\it I^{\pm}}$  correspond to $\eta \pm \chi =
\pi$. One should note that  under the $Z_2$ action, ${\it I^{\pm}}$ are
identified with one another. The conformal group ${{\rm Conf}}(3,1)$ of
Minkowski spacetime  acts smoothly and transitively on compactified identified
Minkowski spacetime $M^{\#}$.
% From
%(2.4), (2.5) and (2.6) it follows that one  may think of $M^{\#}$ as the set
%of null rays in the 6-dimensional flat space with coordinates
%%%$(X^0,X^1,X^2,X^3,X^4,X^5)$ with metric given by
%$$
%ds^2 = (dX^0)^2 + (dX^5)^2 - (dX^1)^2 - (dX^2)^2 - (dX^3)^2 - (dX^4)^2 .
%\eqno (2.18)
%$$
It  now follows easily  from {\mmetric} that ${\rm Conf}(3,1) \equiv SO(4,2)/
\pm 1$.

The conformal compactification can be described in a more group theoretic way
and  will be  useful when discussing the Yang-Mills solutions.
%We begin by recalling the construction of compactified identified Minkowski
%%%spacetime $M^ {\#}$.
This is standard, but the special feature that we shall exploit is perhaps not
as well known as it should be, i.e. the correspondences
\eqn\corr{
M^{\#} \equiv U(2) \equiv U(1) \times SU(2) /Z_2 \equiv S^1 \times S^3 / \pm 1
\equiv S^1 \times S^3 .
}
\noindent To see how {\corr} comes about we identify points in Minkowski
spacetime $M$ with $2 \times 2$
Hermitian matrices $X$:
\eqn\herm{
(t,x^i) \in M \rightarrow X = t + \sigma^i x^i}
  with metric
\eqn\detdx{
ds^2 = -{\rm det} dX.
}
The Cayley map $h: M \rightarrow U(2)$ is given by
\eqn\cayley{
X \rightarrow U = {{1 +iX} \over {1-iX}}
}
and provides a conformal embedding of $M$ into $U(2)$.
% However, $U(2)$ can be regarded as  $ U(1) \times SU(2) /Z_2 \equiv S^1
%%\times %S^3 / \pm 1\equiv M^{\#}$.

  It follows from  {\U} and {\cayley} that
\eqn\detdX{\eqalign{
{\rm det} dX  = \Omega ^{-2} &\biggl (  -(dX^0)^2 - (dX^5)^2 + (dX^1)^2
+(dX^2)^2 +(dX^3)^2 + (dX^4)^2 \biggr )\cr
  &= \Omega ^{-2} \biggl ( -d \eta ^2 +d \chi ^2 + \sin ^2 \chi
( d \theta ^2 + \sin ^2 \theta d \phi ^2) \biggr )\cr
}}
 with
\eqn\omega{
\Omega = \cos \eta + \cos \chi = { 1\over 2} | {\rm det} ( 1 + U)|
}
thus reproducing {\metric}.
The left hand side of {\detdX}  is the standard Minkowski metric, and the right
hand side of  {\detdX} is the standard product metric on the Einstein cylinder
$R \times S^3$ which is the universal covering space $\tilde M ^{\#}$ of
compactified identified Minkowski spacetime $M^{\#}$.  Thus the Cayley
embedding $h$ given by {\cayley} is a conformal one.
The Cayley map $h$ ceases to be invertible when ${\rm det} (1+U)$ vanishes,
i.e. when the conformal factor $\Omega$ vanishes.
These points correspond to null infinity $ I$ (also called \lq \lq scri \rq
\rq). From {\U}, we may express $U$  as
\eqn\stereo{
U = \exp i( \eta + \chi { n }\cdot {\bf \sigma} ),\quad n^i = x^i/r.
}
Note that at $t=0$, (i.e. $\eta =0$) {\stereo} is just the standard
stereographic projection of $S^3$ onto $R^3$.

{\subsec{ Conformally Invariant Equations}}

Given the Yang-Mills solutions on $S^1\times S^3$ and   the conformal
compactification from     Minkowski space to $S^1\times S^3$ discussed above,
we can obtain Yang-Mills solutions in Minkowski space. This makes use of the
conformal invariance of the $3+1$ dimensional Yang-Mills field equations.
When one speaks of conformally invariant equations, one may mean invariance
under the infinite dimensional abelian group of Weyl rescalings of the metric
or the 15 parameter Bateman-Cunningham group ${\rm Conf}(3,1)$
of conformal isometries of Minkowski spacetime  or possibly
just invariance under its 11 dimensional  causality preserving
   Zeeman subgroup
consisting of the semi-direct product of the Poincare group with
spacetime homotheties also called dilatations. This latter group
is sometimes, slightly misleadingly, also called the Weyl group.

Equations which depend only upon
the conformal equivalence class of the spacetime metric $g_{\alpha \beta}$
so that given a solution with respect to  a metric $g_{\alpha \beta}$
it is also (possibly after multiplying by some function
of the conformal factor $\Omega$) a solution  with respect to a conformally
rescaled metric $\Omega^2 g_{\alpha \beta}$ are said to be Weyl invariant.
Important examples are:
 the massless scalar field
\eqn\scalar{
- \nabla ^2 \varphi + { 1 \over 6} R \varphi = 0, \ \ \  \varphi \rightarrow
\varphi /  \Omega,
}
  the massless Dirac equation
\eqn\fermion{
-i \gamma ^ \alpha \nabla _ \alpha \psi =0, \ \ \  \psi \rightarrow \psi /
\Omega ^{ 3 \over 2},
}
  the Yang- Mills equation
\eqn\ym{
g^{\alpha \beta} D _ \alpha F _ {\beta \gamma} =0, \ \ \ F _{\alpha \beta}
\rightarrow F_{\alpha \beta},
}
  and the motion of a perfect radiation fluid
\eqn\fluid{
T^ {\alpha \beta } \  _{; \beta} = 0, \ \ \ u^ \alpha \rightarrow \Omega ^{-1}
u^\alpha , \ \ \rho \rightarrow \Omega ^ {-4} \rho
}
with
\eqn\fluidstress{
T ^ {\alpha \beta } = { 4 \over 3} \rho u ^\alpha u ^ \beta - { 1 \over 3 }
\rho  g ^ {\alpha \beta}
.}
 In all cases the energy momentum tensor $T^{\alpha \beta}$ is traceless
and Weyl rescales as $ T ^ {\alpha \beta} \rightarrow \Omega ^{-6} T ^ {\alpha
\beta}$.

  On the other hand one says that an equation has Bateman-Cunningham invariance
or is ${\rm Conf}(3,1)$ invariant if given a solution $\Phi(x^\alpha)$  then
the pull-back
\eqn\pullback{
J^w \Phi ( c^{-1} x^\alpha)
}
is also a solution.
  $c \in {\rm Conf}(3,1)$ is a conformal isometry of  Minkowski spacetime, $w$
a suitable conformal weight and $J$ the Jacobian of $c$. Every Weyl
invariant equation is ${\rm Conf}(3,1)$ invariant (provided it is
diffeomorphism invariant)
but the converse is not true. Counter-examples are provided by
various higher spin wave equations for zero-mass particles.
Similarly any ${\rm Conf}(3,1)$ invariant equation
is necessarily invariant under its causality preserving subgroup but
again the converse is not true. Essentially any Poincare invariant
equation without
dimension-full coupling constants will be invariant under dilatations.
Examples include a polytropic perfect fluid with a constant ratio of pressure
to density, equations involving the eponymous dilaton, which may be thought of
as a gauge field for dilatations, and the non-linear $\sigma$-model.

A well known strategy for obtaining
solutions of Weyl invariant equations on Minkowski spacetime, and one
which we have  adopted here, is to solve them on the Einstein cylinder,
i.e. on the universal covering space
$\tilde M ^{\#}$ of compactified identified Minkowski spacetime $M^{\#}$
and then to project down onto Minkowski spacetime $M$. Provided the solutions
are well behaved with finite energy (with respect to the Killing vector
${\partial \over {\partial \eta}}$ on $\tilde M ^{\#}$) they will remain
well behaved and of finite energy on  $M$. Note that it is neither
 necessary nor indeed always possible to insist
that the solutions are single valued on compactified identified Minkowski
spacetime $M^{\#}$. This is related in part to the so-called Grgin phenomenon
{\ref\grgin{E. Grgin, {\it J. Math. Phys.} {\bf 9}   (1968) 1595.}}.
Consider for example solutions of the free conformally invariant
wave equation
{\scalar}. One might consider  superpositions of solutions of the form:
\eqn\harm{
\varphi = \exp (i( l+1) \eta ) \ Y_l ( \chi, \theta, \phi)
}
where $Y_l( \chi, \theta, \phi)$ is a hyper-spherical harmonic of degree $l$
on the 3-sphere $S^3$.
To obtain $M^{\#}$, points $(\eta, \chi, \theta, \phi)$ and $(\eta + \pi ,\pi -
\chi, \pi -\theta, \phi + \pi)$ must be identified, but from {\harm}
and the properties of harmonics
\eqn\phitrans{
\varphi (\eta, \chi, \theta, \phi)= -\varphi (\eta + \pi ,\pi - \chi, \pi
-\theta, \phi + \pi) .
}
Thus solutions $\varphi$  of the form {\harm} are not single-valued, in fact
they may be thought of as sections of a flat bundle over $M^{\#}$ twisted by a
representation of the conformal group ${\rm Conf}(3,1)$ in $\pm 1$
{\ref\lerner{D. Lerner, {\it Twistor Newsletter}  {\bf 3} (1976) 7;
N. Woodhouse,  {\it Twistor Newsletter} {\bf 6} (1977) 1;
C. Clarke and D. Lerner, {\it Comm. Math. Phys.} {\bf 55} (1977) 179.}}.
This is consistent with the fact that $\varphi$ scales as one power of $\Omega$
and that if one extends $\Omega$ smoothly over
the Einstein cylinder $R \times S^3$ using {\metric}
it is anti-periodic under the identification.

  Physically, the solutions {\harm} may be thought of as an incoming wave which
starts off from
past null infinity ${ I}^-$, i.e. from $\eta - \chi = - \pi$
comes to a focus at the origin of spherical polars, (i.e. $r=0$) re-expands and
reaches future null infinity ${ I}^+$, i.e. $\eta + \chi = \pi$. Because it has
passed once through a spherical focus or caustic, its amplitude is, by the
well-known G\"uoy effect {\ref\guoy{ G. G\"uoy, {\it C. R. Acad. Sci. Paris}
{\bf 110}   (1890) 125.}}, multiplied by minus one. Note that ${ I}^-$ and ${
I}^+$ are identified in $M^{\#}$.
 %It will be  interesting to compare this behaviour with the collapse of a
%%%cosmic textures in section 6.

\subsec {Yang-Mills Solutions in Minkowski Space}

We now obtain the Yang-Mills solutions on Minkowski space by projecting
the   $SO(2) \times SO(4)$ invariant $f=1/2$ solution   {\conn} from $M^{\#}$.
Note that by contrast with the general $SO(4)$ invariant solution, the $SO(2)
\times SO(4)$ invariant solution
is in fact well defined on identified compactified Minkowski spacetime
$M^{\#}$. Substituting $g = \exp i  \chi { n }\cdot { \sigma} $ from {\stereo}
into {\conn}   yields
\eqn\gaugefield{\eqalign{
A_u^a & = {2\over er}({n^a\over 1+u^2}),\quad u =t-r\cr
A_v^a &= -{2\over er}({n^a\over 1+v^2}),\quad v =t+r\cr
A_i^a  =   - {2\over e(1+u^2) (1+v^2)}& \biggl (2\epsilon_{iaj}x^j  +
(1+uv)P^{ia} \biggr ),
\quad \quad P^{ij} = {\delta^{ij} - n^in^j} .\cr
} }
The components of the gauge field are peaked on the future and past light cone
$u=v=0$. The stress-energy tensor for the    gauge field can be obtained
from the stress tensor of the gauge field on ESU. Under a conformal
transformation $\tilde g_{\mu\nu} = \Omega^2 g_{\mu\nu}$, the energy density
transforms as $\tilde\rho = \Omega^{-4}\rho$.        Weyl rescaling the energy
density {\energydensity}
of the solution on ESU to Minkowski spacetime $M$,
we obtain the  energy  density (using {\energydensity} and {\metric}))
\eqn\rhotilde{
\tilde \rho = {{192E} \over { e^2}} {1 \over { ( 1+u^2)^2 (1+v^2)^2}}
}
which is indeed  peaked on the past and future light cone. The Chern-Simons
number $N_{CS}$
of the solution evaluated on any Cauchy surface in Minkowski spacetime $M$
equals $1 \over 2$. Hence, by itself this solution does not lead to a change
of Chern-Simons number. Indeed, we are envisioning this solution
as an approximation to the latter part of a process in which one
begins with the static sphaleron configuration.
Solutions to the pure Yang-Mills equations which do involve a change of
Chern-Simons number were discussed in {\ref\farhi{E. Farhi, V. Khoze, and R.
Singleton,
{\it Phys. Rev. D} {\bf 47} (1993) 5551; E. Farhi,  V. Khoze, and R.
Singleton,
{\it Phys. Rev. D} {\bf 50} (1994) 4162.}}.
Thus we have a  time-symmetric shell (moving at the speed of light) which
carries in an amount $N_{CS}= { 1\over 2}$ from past null infinity ${ I} ^-$
and carries out the same amount
back to future null infinity ${ I} ^+$. The solution has a moment of time
symmetry, i.e. it is momentarily at rest on the spacelike hyperplane $t=0$.
The total energy of the shell carrying this Chern-Simons number
is finite because the field falls
off rapidly near infinity. The flow vector $u^ \alpha$ in Minkowski spacetime
will be orthogonal to the surfaces
$\eta = {\rm constant}$, i.e. to the surfaces
\eqn\taneta{
\tan \eta = {{u+v} \over {1-uv} } = {{2t} \over {1 + r^2 - t^2 }} = {\rm
constant}
}
and is given by
\eqn\umu{
u^{\mu} = \Omega \biggl ({\partial\;\over \partial\eta}\biggr )^{\mu}=  \biggl
({ 1+u^2\over 1+v^2}\biggr )^{1/2} \biggl ({\partial\;\over \partial u}\biggr
)^{\mu}  +  \biggl ({ 1+v^2\over 1+u^2}\biggr )^{1/2}\biggl ({\partial\;\over
\partial v}\biggr )^{\mu}
.}

{\newsec{Other Gauge Groups}}

It is possible to consider $SO(4)$ invariant solutions for
other gauge groups $G$. Bertolami, et. al. {\ref\bert{ O. Bertolami, J. M.
Mouro, R. F. Picken, I. P. Volobujev, {\it Int. J. Mod. Phys.} {\bf A6} (1991)
4149.} study  the gauge group $G= SO(n)$.  Setting $A_0=0$, then their gauge
field can be expressed in terms of the time-dependent functions
$(\chi_0, \chi_i),\;i=1,\dots n-3 .$ $\chi_0$ corresponds to the $SU(2)$
subgroup  and is related to our $f$ by $\chi_0 =2f-1$.     Like the
$SU(2)$ solutions, these $ SO(n)$ solutions can be expressed in terms of a
particle moving in the potential:
\eqn\potential{
V (\chi_0, \chi_i) ={ 1 \over 8} \left ( ( 1 -\chi _0 ^2 - \chi ^2 ) ^ 2 + 4
\chi _0 ^ 2 \chi ^2 \right ),\quad \chi^2 = \chi^i\chi^i.
}
The associated Chern-Simons number (in their  normalization) is given by
\eqn\chernsimons{
N_{CS} = 2 + 3 \chi_0 - \chi_0 ^3 - 3 \chi _0 \chi ^2 .
}
 For $\chi^i=0$,   {\redeq} and {\cs} are recovered (taking into account
a difference of normalization).   These solutions can be mapped to solutions in
Minkowski space which have  qualitatively similiar properties to the  $SU(2)$
cases. The energy density of the resulting gauge field will have  the same
form, but with a different numerical factor which can be determined from
{\potential}.
There are four vacua on $\tilde {\cal C}$   having zero energy and
which are absolute minima.
These are at $(\chi_0, \chi) = (1,0), (-1,0), (0,1), (0,-1)$ with $N_{CS}=
4,0,2,2$ respectively.
There is a local maximum
at $(0,0)$ with energy $1 \over 8$ and  $N_{CS}=2$ and four
saddle points $({1 \over 2},{ 1\over 2}), ({ 1\over 2}, -{ 1\over 2} ),
( - { 1 \over 2} ,{ 1 \over 2}), (- { 1\over 2}, - { 1\over 2} )$
 having energy
$1 \over {16}$ and
   $N_{CS} = 3,3,1,1$.
It is interesting to note the existence of  sphaleron
configurations with lower energy than the pure $SU(2)$ one
that
we have considered above.

{\newsec{Spinors}}

It is straightforward to solve the massless
Dirac equation on the Einstein cylinder in
the background of the  $SO(2) \times SO(4)$ invariant $f= {1 \over 2}$ solution
 {\us}{\ref\ho{ A. Hosoya and W. Ogura, {\it Phys. Lett. } {\bf B225}
(1989) 117.}}}. In particular one can find
time-independent (i.e. $\eta$ independent)
zero energy  solutions of the coupled Dirac equation on the Einstein cylinder.
Scalar quantities such as ${\overline \psi} \psi$ are homogeneous for
the zero-energy solutions.
These solutions may be Weyl rescaled to give time dependent
solutions of the Dirac equation in the time-dependent sphaleron.
{}From {\fermion}, scalar quantities such as ${\overline {\tilde \psi}} {\tilde
\psi}$ will now be proportional to $\Omega ^3$ and will therefore be small
except where
$\Omega$ is large. In other words these special solutions of the Dirac equation
will have their support  peaked  on the shell and may be thought of as carried
by the shell.

\bigbreak\bigskip\bigskip\centerline{\bf Acknowledgements}\nobreak
   A.S.
was supported in part by  the SERC at Cambridge and
NSF grant NSF-PHY-93-57203 at Davis.
\baselineskip=30pt

\footatend\vfill\supereject\immediate\closeout\rfile\writestoppt
\baselineskip=14pt\centerline{{\bf References}}\bigskip{\frenchspacing%
\parindent=20pt\escapechar=` \input refs.tmp\vfill\eject}\nonfrenchspacing
\end